\documentclass[iop, jmp, amssymb, reprint, graphicx]{revtex4-1}

\usepackage{braket}

\usepackage{color}

\newcommand{\lp}{\left(}
\newcommand{\rp}{\right)}

\newcommand{\lsb}{\left[}
\newcommand{\rsb}{\right]}

\newcommand{\refeq}[1]{Eq. (\ref{#1})}
\newcommand{\refsec}[1]{Sec. \ref{#1}}
\newcommand{\refapp}[1]{Appendix \ref{#1}}
\newcommand{\reffig}[1]{Figure \ref{#1}}

\newcommand{\grad}{\nabla}

\newcommand{\KAISTNQe}{Department of Nuclear and Quantum Engineering, KAIST, Daejeon 34141, Korea}

\newcommand{\NFRI}{National Fusion Research Institute, Daejeon, Republic of Korea}

\newcommand{\jwkim}{\author{Jaewook~Kim}\email[]{ijwkim@kaist.ac.kr}\affiliation{\KAISTNQe}}
\newcommand{\hhkhang}{\author{H.H.~Khang}\affiliation{\NFRI}}
\newcommand{\jhogun}{\author{Hogun~Jhang}\affiliation{\NFRI}}
\newcommand{\sskim}{\author{S.S.~Kim}\affiliation{\NFRI}}
\newcommand{\wjlee}{\author{Wonjun~Lee}\affiliation{\KAISTNQe}}

\newcommand{\ycghim}{\author{Y.-c.~Ghim}\email[]{ycghim@kaist.ac.kr}\affiliation{\KAISTNQe}}

\usepackage{amsmath}
\usepackage{graphicx}
\usepackage{epstopdf}
\usepackage{dcolumn}
\usepackage{bm}
\usepackage{chngcntr}
\usepackage{footmisc}

\draft 
\usepackage[english]{babel}

\newcounter{subeqn} %
\makeatletter
\@addtoreset{subeqn}{equation}
\makeatother

\begin{document}

\title{Evolution of magnetic Kubo number of stochastic magnetic fields during the edge pedestal collapse simulation}

\jwkim 
\wjlee
\jhogun
\sskim
\hhkhang
\ycghim

\date{\today}
\begin{abstract}
Using a statistical correlation analysis, we compute evolution of the magnetic Kubo number during an edge pedestal collapse in nonlinear reduced magnetohydrodynamic simulations. The kubo number is found not to exceed the unity in spite of performing the simulation with a highly unstable initial pressure profile to the ideal ballooning mode. During the edge pedestal collapse, the Kubo number is within the values of $0.2$ and $0.6$ suggesting that the quasilinear diffusion model is sufficient to explain the energy loss mechanism during the pedestal collapse. Temporal evolution of poloidal correlation lengths of pressure fluctuations resembles with that of the Chirikov parameter and the Kubo number; while radial correlation lengths of the pressure fluctuations are strongly correlated with the radial width of the magnetic stochastic layer.
\end{abstract}

\maketitle

\section{introduction}
\label{sec:intro} 

The abrupt edge pedestal collapse due to the onset of type-I edge localized modes (ELMs) must be mitigated or suppressed as they can damage plasma facing components in reactor-grade fusion experiments. During the last decade, experimental techniques to mitigate\cite{Evans:2013hzba,Suttrop:2011ksba,Suttrop:2011cyba} and/or suppress\cite{Jeon:2012khba, Evans:2005ecba, Evans:2006hrbaca, Burrell:2005hr, Evans:2008dmba} ELMs have been developed. Together with such a progress in experiments, many theoretical works have been reported. Regarding the onset of type-I ELMs, the popular idea is based on the destabilization of the linear ideal peeling ballooning modes.\cite{Snyder:2002fsba, Snyder:2009hgba} Recently, nonlinear magnetohydrodynamics (MHD) simulations\cite{Dudson:2011foba,Huysmans:2009dqba,Sugiyama:2010baba} have been carried out to enhance our physical understanding of the ELM crash and the subsequent energy loss mechanism.

A recent study carried out at MAST (Mega-Amp Spherical Tokamak) suggested that energy (particle) loss through the filament structures\cite{Kirk:2006jp,Scannell:2007ecba} accounts for only $15\:(25)$\:\% of the total loss during an ELM crash.\cite{Kirk:2007doba} A question then arises as to which process can account for the major part of the energy release. Many nonlinear simulations such as JOREK,\cite{Huysmans:2009dqba} BOUT++\cite{Xu:2010bnba, Rhee:2015czbaca, Jhang:2017jm} and M3D\cite{Sugiyama:2012bqba,Sugiyama:2010baba} indicate that stochastic (or ergodized) magnetic fields are generated during an abrupt edge pedestal collapse. Quite naturally, one is led to consider the possibility of the energy loss through the stochastic magnetic fields.\cite{Rechester:1978cl}

One way to identify a transport mechanism due to stochastic magnetic fields is based on the magnetic Kubo number\cite{Zimbardo:2000fm,Zimbardo:2000cv} defined as
\begin{equation}
\label{eq:kubo_number}
\mathcal{R}= \frac{ \delta B }{B_0} \frac{l_\parallel}{l_\perp},
\end{equation}
where $\delta B$ and $B_0$ are magnitudes of perturbed and equilibrium magnetic fields, respectively. $l_\parallel$ and $l_\perp$ are the correlation lengths of parallel and perpendicular, with respect to the equilibrium magnetic field, components of the perturbed magnetic fields, respectively. For $0.2\lesssim \mathcal{R} \lesssim 1.0$, cross-field transport approximately follows quasilinear Gaussian diffusion, whereas the percolation theory must be used if $\mathcal{R} \gg 1 (\mathcal{R} \gtrsim 10)$.\cite{Kadomtsev:1979,Zimbardo:2000fm,Zimbardo:2000cv} If $\mathcal{R}\lesssim 0.2$, then turbulence driven anomalous transport dominates the cross-field diffusion.\cite{Zimbardo:2000fm,Zimbardo:2000cv}

In this work we carry out a correlation analysis with the aim of investigating how the magnetic Kubo number evolves during and after an abrupt edge pedestal collapse, thereby we may suggest a transport mechanism during the collapse. We use the simulation results reported in \citet{Rhee:2015czbaca} and \citet{Jhang:2017jm}. Based on our analysis, $\mathcal{R}$ is found to be within the values of $0.2$ and $0.6$ during the collapse, and becomes less than $0.1$ as the collapse is completed. This result suggests that the percolation theory may not be necessary to describe the process of ELM-induced energy release. 

A direct experimental validation of our work is hindered due to lack of diagnostic systems capable of measuring spatially localized fine-scale (possibly ion-scale) stochastic magnetic fields. We suggest that temporal evolutions of poloidal and radial correlation lengths of pressure fluctuations may be used as circumstantial evidences of the existence of stochastic magnetic fields and the evolution of $\mathcal{R}$. We find that the evolution of the poloidal correlation length resembles strikingly close to that of both Kubo number and Chirikov parameter, while the radial correlation length is strongly correlated with the radial width of the stochastic layer.

We first describe our numerical model and present features of the pedestal collapse simulation with general results in \refsec{sec:nume_model}. Then, we discuss how various correlation lengths of perturbed magnetic fields as well as the estimated Kubo number evolve during and after an abrupt edge pedestal collapse in \refsec{sec:mag_corr_length_section} and \refsec{sec:kubo_number} followed by behavior of correlation lengths of pressure fluctuations. For the readers who are interested in how we have estimated various statistical quantities such as parallel, poloidal and radial correlation lengths of perturbed quantities, we provide a detailed description of the technique we have employed in this work in \refsec{sec:mag_corr_length_section} and \refapp{app:interpolation_method}. Our conclusion is stated in \refsec{sec:conclusion}.

\section{Simulation model}
\label{sec:nume_model}
The main objective of this paper is to evaluate various correlation lengths (parallel, radial and poloidal) of stochastic magnetic fields and the magnetic Kubo number using the simulation resutls reported in \citet{Rhee:2015czbaca} and \citet{Jhang:2017jm}. For completeness, we provide a brief summary of the numerical model and main results of the simulation.

\subsection{Numerical model}
\label{subsec:model}

The edge pedestal collapse simulations have been performed using a three-field reduced MHD model. Evolution equations for these three fields are given by,
\begin{eqnarray}
\label{eq:vorticity}
m_i n \left[ \frac{\partial}{\partial t} + \vec{V}_E\cdot\grad \right] U & = & B_0^2 \grad_\parallel \left(\frac{J_\parallel}{B_0}\right)+2\vec{b}_0\times\vec{\kappa}_0\cdot\grad P,   \nonumber \\
\end{eqnarray}

\begin{eqnarray}
\label{eq:pressure}
\left[ \frac{\partial}{\partial t} + \vec{V}_E\cdot\grad \right] P & = & f_K v_e^2 D_{RR} \frac{\partial^2 P_0}{\partial r^2} \nonumber \\ 
& & - \frac{10}{3} \frac{P_0}{1+5\beta_0 /6} \frac{\vec{b}_0 \times \vec{\kappa}_0 \cdot \grad \Phi}{B_0},
\end{eqnarray}
and
\begin{equation}
\label{eq:ohm}
\frac{\partial A_\parallel}{\partial t}=-\grad_\parallel\Phi + \frac{\eta}{\mu_0}\grad_\perp^2 A_\parallel - \frac{\eta_H}{\mu_0}\grad_\perp^4 A_\parallel,
\end{equation}
where $U=\left(1/B_0)(\grad^2_\perp\Phi+(1/en)\grad^2_\perp P\right)$ is the vorticity, $P$ the pressure and $A_\parallel$ the vector potential. $B_0$ is the magnitude of the equilibrium magnetic field, $\Phi$ the total electric static potential, $n$ the plasma density, $\vec{V}_E=\vec{b}_0\times \grad\Phi/B_0$, $\vec{b}_0=\vec{B}_0/B_0$, and $J_\parallel=J_{\parallel 0}-\grad_\perp^2 A_\parallel/\mu_0$ is the total (both equilibrium and perturbed) parallel current density with the vacuum permeability $\mu_0$. $e$ is the charge of a proton, $m_i$ the ion mass, $\eta$ the resistivity, and $\vec{\kappa}_0=\vec{b}_0\cdot\grad \vec{b}_0$ and $\beta_0=2\mu_0 P_0/B_0^2$ denote the magnetic curvature vector and the normalized plasma pressure, respectively. The subscript $0$ ($1$) indicates the equilibrium (perturbed) quantity, while a quantity without a subscript means the total quantity, i.e., equilibrium plus perturbed.

All simulations are performed using the BOUT++ framework\cite{Dudson:2009ig} with the following equilibrium parameters: $R_0=3.5$\;m (major radius), $a=1.0$\:m (minor radius), $B_0=1.94$\:T at the magnetic axis, the Lundquist number $S=\mu_0 R_0 V_A/\eta = 10^9$, and the hyper-Lundquist number $S_H =\mu_0 R_0^3 V_A/\eta_H = 10^{12}$. Here, $V_A$ is the Alfv\'{e}n speed. The role of $\eta_H$ is extensively discussed elsewhere.\cite{Xu:2010bnba, Xu:2011egba}

The first term of the right-hand side (RHS) in the pressure evolution equation, \refeq{eq:pressure}, acts as a parallel heat loss due to stochastic magnetic fields based on the Rechester-Rosenbluth model\cite{Rechester:1978cl} taking a kinetic effect into account.\cite{Park:2010klba} The second term of the RHS in \refeq{eq:pressure} is introduced to include the compressibility effect of $E\times B$ drift\cite{Jhang:2017jm} and is responsible for the GAM (geodesic acoustic mode) generation via a linear coupling between $\Phi_{m=0,\:n=0}$ and $P_{m=1,\:n=0}$ discussed elsewhere,\cite{Diamond:2005etba} where $m$ and $n$ denote the poloidal and toroidal mode numbers, respectively. As we will evaluate the magnetic Kubo numbers with and without zonal flows, the difference between the two is whether or not to include this term, i.e., simulation result with the zonal flow includes this term.

\subsection{Simulation results}
\label{subsec:sim_results}

The simulation initiates with a strong equilibrium pressure gradient and an initial perturbation on the vorticity $U$. We set the value of the normalized pressure profile defined as $\alpha=-2\mu_0 q^2 R_0 (dP_0/dr)/B_0^2$ to be 3.87 at the location of the largest pressure gradient ($R=4.59$\:m). Here, $q$ is the safety factor. This value is much greater than the critical value $\alpha_c=2.75$ beyond which the ideal ballooning mode becomes unstable.\cite{Connor:1978jsba, Connor:1998huba} Consequently, the initially seeded most unstable ballooning mode $n=20$ grows exponentially during the linear state, develops a series of tearing mode activities via nonlinear interactions, and eventually creates stochastic magnetic fields with a rapid collapse of the edge pedestal.\cite{Rhee:2015czbaca}

\begin{figure}[t]
\includegraphics[width=\linewidth]{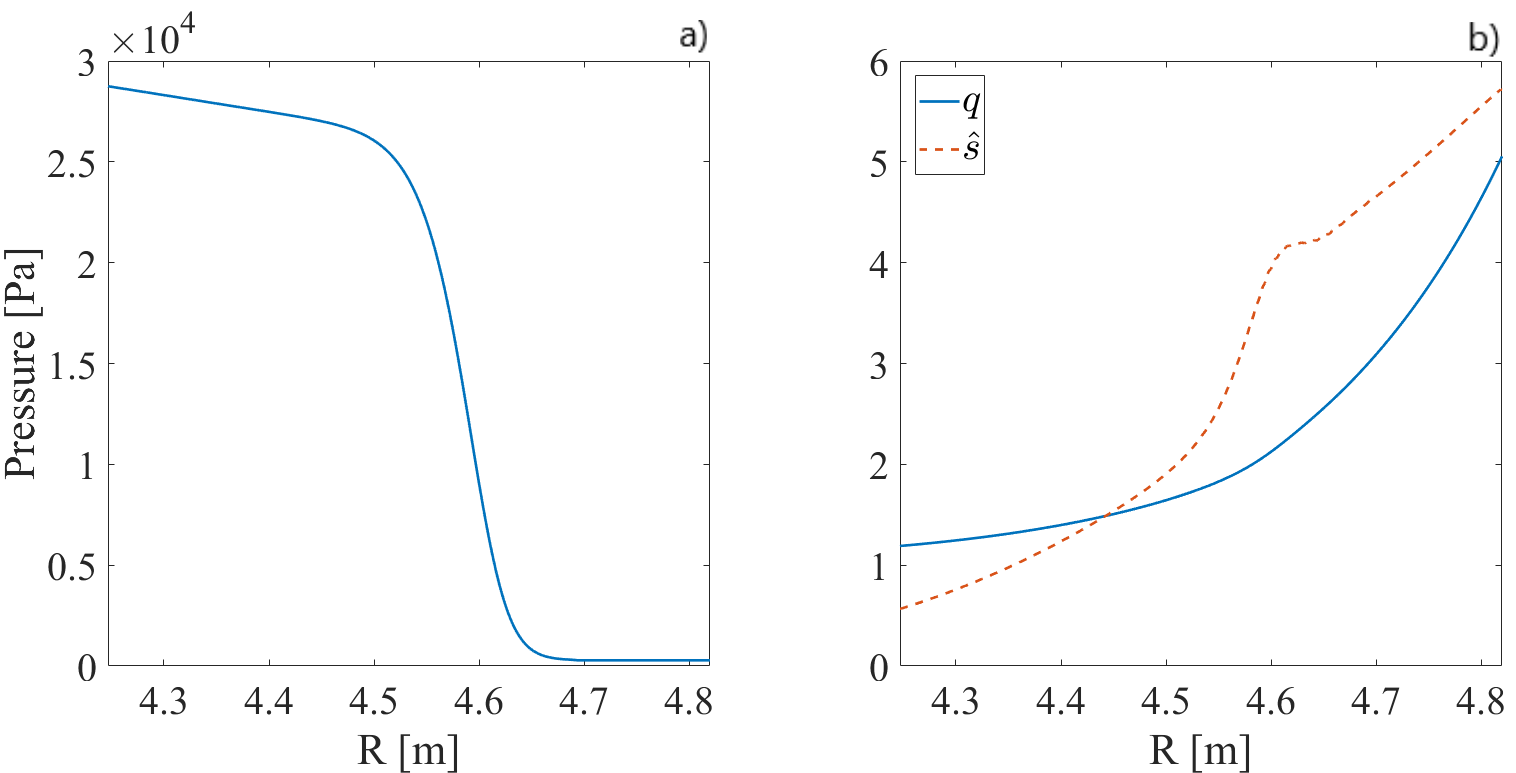}
\caption{Initial profiles of (a) the pressure and (b) the safety factor $q$ (blue line) with the magnetic shear $\hat s$ (red dashed line) at the low field side of a model tokamak being used in simulations.}
\label{fig:initial_profile}
\end{figure}

\begin{figure}[t]
\includegraphics[width=\linewidth]{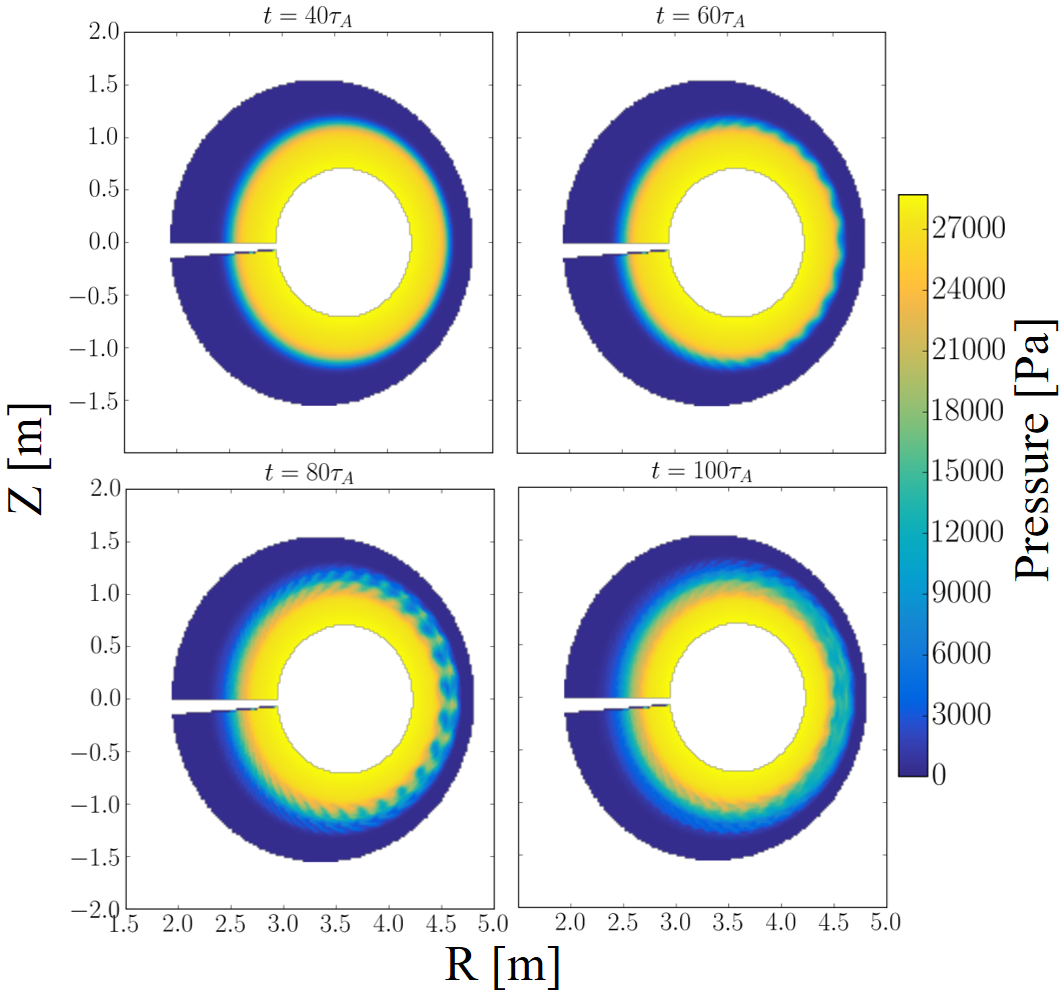}
\caption{Poloidal cross-sections of total pressure contours at $t=40$, $60$, $80$, $100\:\tau_A$.}
\label{fig:pressure_contour}
\end{figure}

\reffig{fig:initial_profile} shows the initial profiles of (a) the pressure and (b) the safety factor $q$ (blue line) with the magnetic shear $\hat s=(r/q)(dq/dr)$ (red dashed line) used in the simulation. \reffig{fig:pressure_contour} shows poloidal cross-sections of total pressure contours at different simulation times, i.e., $t=40$, $60$, $80$, $100\:\tau_A$, where $\tau_A$ is the Alfv\'{e}n time. 

\begin{figure}[t]
\includegraphics[width=\linewidth]{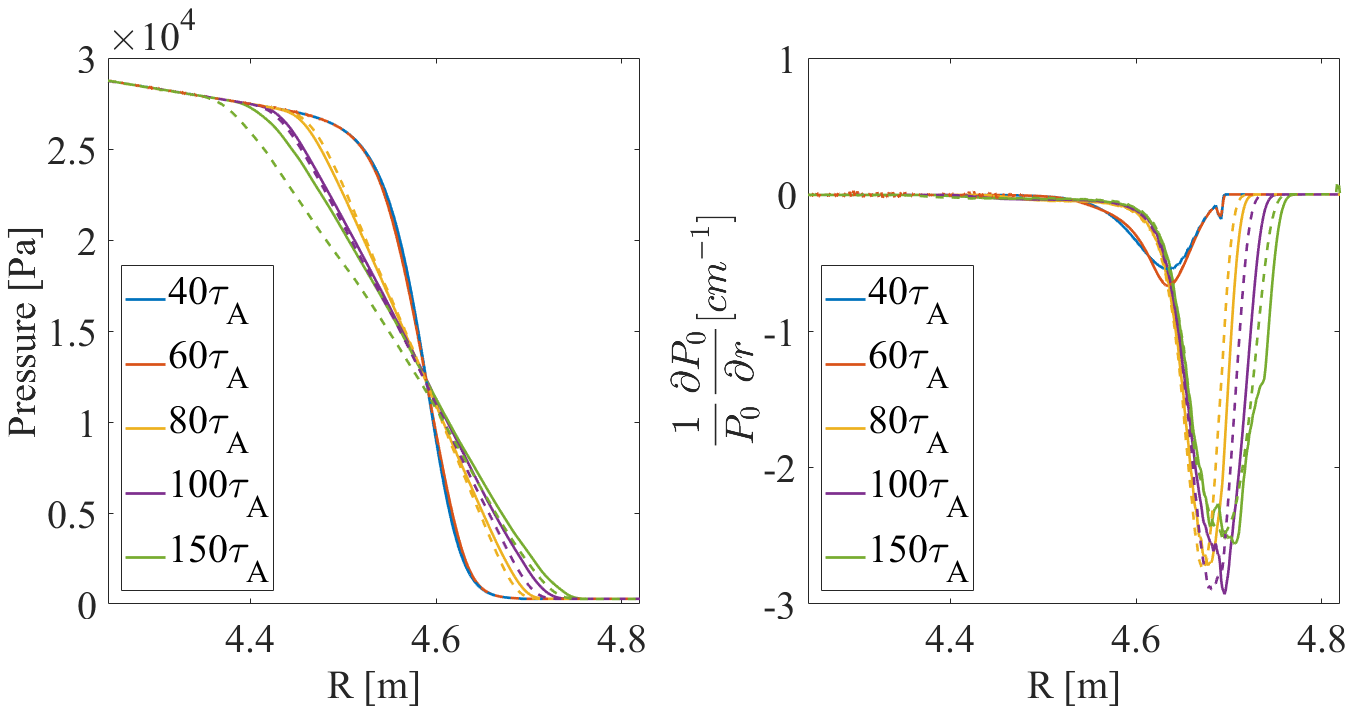}
\caption{(a) Equilibrium pressure profiles and (b) normalized pressure gradient, i.e., $\grad\ln P_0$ at $t=40$, $60$, $80$, $100$, $150\:\tau_A$ with (dashed lines) and without (lines) zonal flows. An abrupt edge pedestal collapse occurs in between $60<\tau_A<80$.}
\label{fig:pressure_and_gradient_profile}
\end{figure}

\reffig{fig:pressure_and_gradient_profile} shows the evolutions of (a) the equilibrium pressure profile and (b) the normalized pressure gradient, i.e., $\grad \ln P_0$, with (dashed lines) and without (solid lines) zonal flows at $t=40$, $60$, $80$, $100$, $150\:\tau_A$. Pressure profiles at $t=40$, $60\:\tau_A$ are quite similar, while an abrupt edge pedestal collapse occurs in between $60<\tau_A<80$. We soon show that the non-linear state develops around $\tau_A\approx 60$. Note that for $t=40$ and $60\:\tau_A$, solid and dashed lines are indistinguishable as they overlap each other in the figure. 

Since we are interested in obtaining the magnetic Kubo number based on the correlation lengths of perturbed magnetic fields, we need to distinguish between the linear and non-linear states within the simulation duration. We note that the estimated correlation lengths from the linear state do not capture a significant physical meaning, i.e., wavenumber, frequency and/or phase information are more relevant for the linear state, because correlation is, by definition, \textit{statistically} estimated as described in \refsec{sec:mag_corr_length_section}.

During the linear state any perturbed (scalar) physical quantities as a function of space $x$ and time $t$ denoted as $\xi(x, t)$ can be expressed as  
\begin{equation}
\xi(x, t) \sim e^{i\left(kx-\omega t + \delta \right)} e^{\gamma t},
\end{equation}
where $k$, $\omega$ and $\delta$ are the wavenumber, the frequency and the phase of the perturbed quantity $\xi$, respectively, with the linear growth (or damping) rate of $\gamma$. Owing to the fact that $\xi(x, t)$ in our simulation has a well defined wave-like structure on a magnetic flux surface, the standard deviation of $\xi$ along the poloidal direction within a flux surface can be used as a fluctuation level of $\xi$ denoted as $||\xi ||$:
\begin{equation}
||\xi ||(t) = \frac{1}{2\pi}\left[ \int_0^{2\pi}\:dx\:\left\{\text{Re}\left[\xi(x,t)\right] \right\}^2   \right]^{1/2}\sim e^{\gamma t},
\end{equation}
where $\text{Re(\:)}$ returns the real part of the argument, and $x$ is taken as the poloidal direction here. We expect that the quantity $\xi(x_i, t)/||\xi ||(t) \sim e^{i\left(kx_i-\omega t + \delta \right)}$ at a fixed spatial position of $x=x_i$ during the linear state, resulting in a clear wave-like feature as a function of time. 

\begin{figure}[t]
\includegraphics[width=\linewidth]{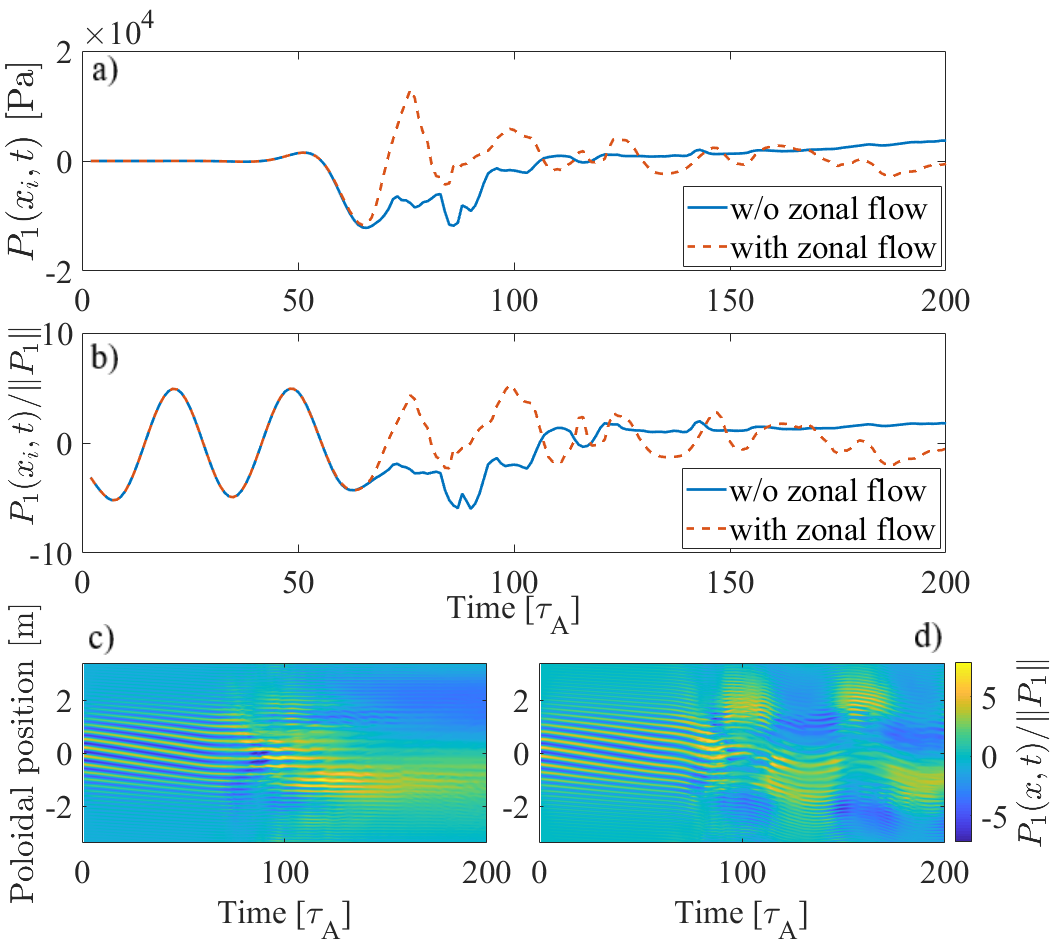}
\caption{(a) $P_1(x_i, t)$ with (red dashed line) and without (blue line) zonal flows at a fixed spatial position $x=x_i$; while (b) $P_1(x_i, t)/||P_1||(t)$ clearly showing wave-like structure when $\tau_A \lesssim 60$. (c) and (d) are the contours of the same quantity as in (b) at all poloidal positions on a flux surface with (d) and without (c) zonal flows. As we expect the clear wave-like structure during the linear state, $\tau_A \lesssim 60$ corresponds to the linear state, and the non-linear state is followed in all cases. (c) and (d) also show advection of $m=40$ ballooning mode during the linear state due to the diamagnetic drift.}
\label{fig:fluctuation_divided}
\end{figure}

\reffig{fig:fluctuation_divided}(a) shows $\xi(x_i, t)$ where the perturbed pressure $P_1$ is used as $\xi$ with (red dashed line) and without (blue line) zonal  flows. As the level of perturbation grows, it hardly shows any well defined wave-structures; while it is evident in (b) showing $\xi(x_i, t)/||\xi ||(t)$ that $\tau_A \lesssim 60$ is the linear state followed by the non-linear state later on. Thus, our method is useful to distinguish between linear and non-linear states. \reffig{fig:fluctuation_divided}(c) and (d) are the contours of the same quantity as in (b) for all poloidal positions on a flux surface. They do also show a clear development of the non-linear state approximately after $\tau_A \approx 60$. Furthermore, these contours manifest poloidal rotation of the $m=40$ ballooning mode during the linear state, and its phase speed is found to be $14$\:km/s using the cross-correlation time delay method,\cite{Ghim:2012keba} where the diamagnetic drift velocity due to the pressure gradient is estimated to be $18$\:km/s. Our simulation during the linear state is consistent with the physical picture of the linear ballooning mode.\cite{Morales:2016dqbaca}

\begin{figure}[t]
\includegraphics[width=\linewidth]{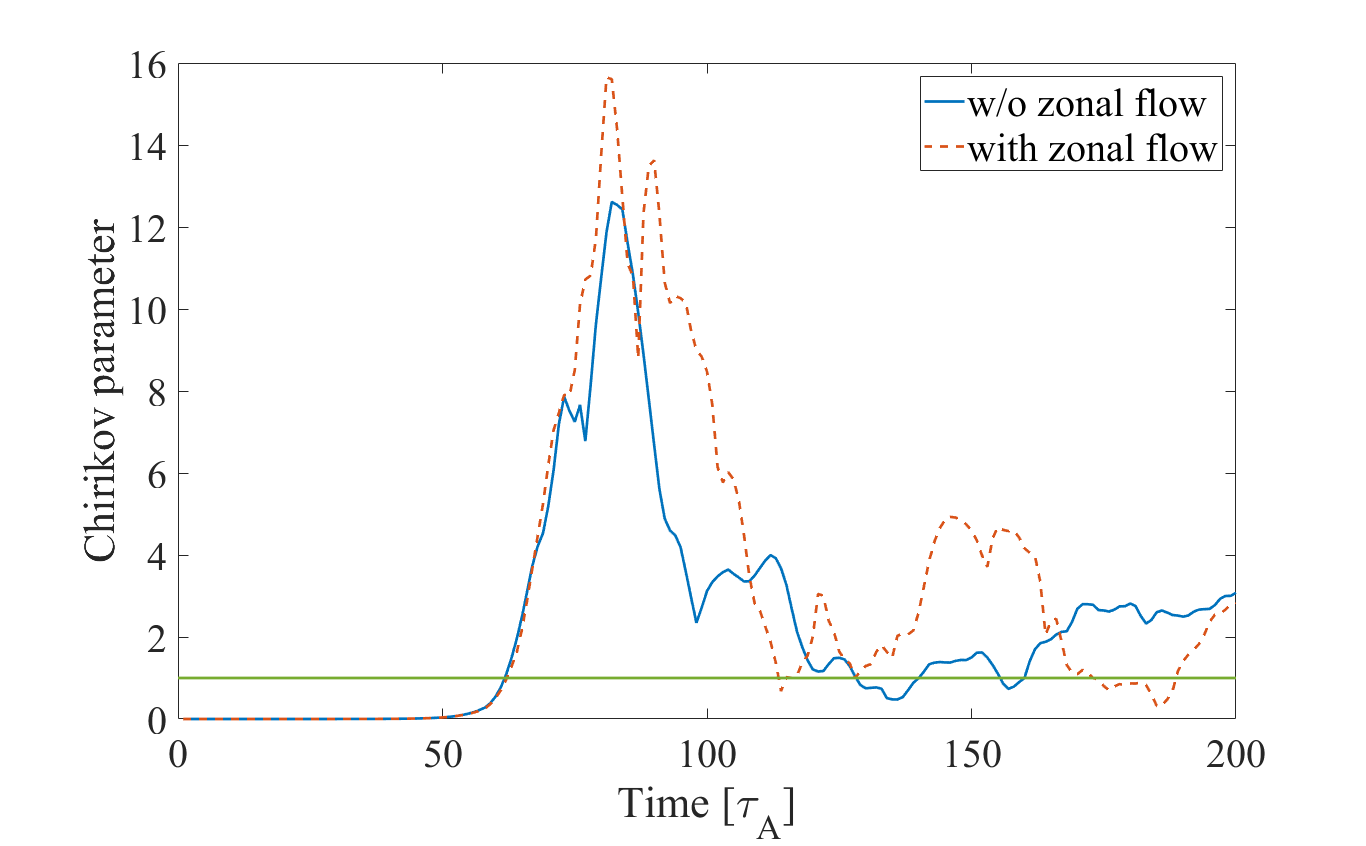}
\caption{Evolutions of the Chirikov parameter $\mathcal{C}$ at the location of the maximum pressure gradient with (red dashed line) and without (blue line) zonal flows. $\mathcal{C}>1$ during $\tau_A\gtrsim 60$, indicating stochastization of the magnetic fields. The green horizontal line indicates $\mathcal{C}=1$.}
\label{fig:chirikov_parameter}
\end{figure}

As shown in \reffig{fig:pressure_and_gradient_profile}(a), the pressure profile hardly changes during $\tau_A \lesssim 60$, while an abrupt pedestal collapse is observed after $\tau_A \approx 60$. This also indicates that the development of the non-linear state during $\tau_A \gtrsim 60$, causing the stochastization of the magnetic field and a subsequent sudden increase of cross-field diffusion. For this reason, we also estimate the Chirikov parameter\cite{Chirikov1979,Rechester:1978cl, Biewer:2003dnba} denoted as $\mathcal{C}$:
\begin{equation}
\mathcal{C}=\frac{W_m+W_{m+1}}{\Delta\psi_{m,m+1}},
\end{equation}
where $W_{m\:(m+1)}$ is the size of magnetic islands for the mode number $m\:(m+1)$, and $\Delta\psi_{m, m+1}$ is the distance between the two neighboring rational surfaces. If $\mathcal{C}>1$, then the neighboring magnetic islands overlap and the magnetic field stochastization occurs. We show the estimated (at the location of the maximum pressure gradient) Chirikov parameter $\mathcal{C}$ in \reffig{fig:chirikov_parameter} both with (red dashed line) and without (blue line) zonal flows. It is clear that $\mathcal{C}$ becomes larger than one (green horizontal line) around $\tau_A \approx 60$ corresponding to the time entering the non-linear state. 

\begin{figure}[t]
\includegraphics[width=\linewidth]{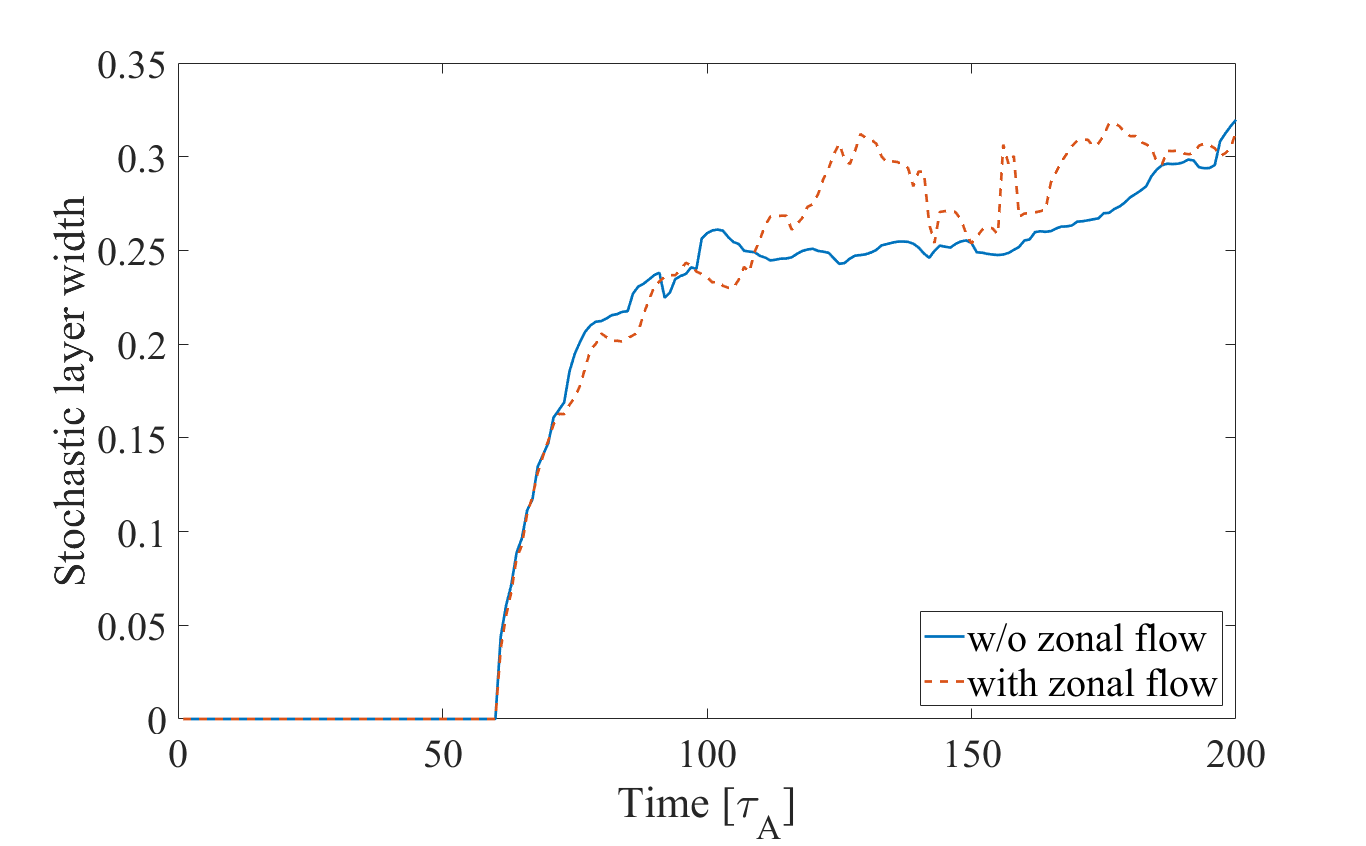}
\caption{Evolutions of stochastic layer width in radial direction with (red dashed line) and without (blue line) zonal flows.}
\label{fig:stochastic_layer}
\end{figure}

\reffig{fig:stochastic_layer} shows the evolutions of widths (in radial direction) of the observed stochastic layers with (red dashed line) and without (blue line) zonal flows showing a rapid increase of the layer widths while $\mathcal{C}$ increases abruptly, i.e., $60\lesssim \tau_A \lesssim 80$, then the layer widths slowly increase as $\mathcal{C}$ falls down to a plateau region, i.e., $\tau_A \gtrsim 130$. Note that we have set the stochastic layer width to zero for $\tau_A < 60$ since there is no stochastic magnetic field during the linear state as attested by $\mathcal{C}$ in \reffig{fig:chirikov_parameter}, i.e., $\mathcal{C}<1$ for $\tau_A<60$.

\section{Correlation lengths of stochastic magnetic fields}
\label{sec:mag_corr_length_section}

Experimental measurements of correlation lengths in fusion-grade hot plasmas are typically performed by replacing the ensemble average operator with a temporal-domain average operator assuming that the data are ergodic, stationary and homogeneous.\cite{randomdata, Kim:2016juba, Kim:2016kdba} Since we only have $201$ points in time in the simulation data, i.e., $\tau_A=[0, 1, 2, ..., 199, 200]$, the temporal-domain average is not suitable to obtain statistically valid results. Furthermore, running tens of the same simulation to obtain the ensemble average is computationally too expensive. 

Therefore, we perform an ensemble average by obtaining data within a flux surface but following many different field lines with separation distance of $\sim 44$\:cm between the neighboring field lines in toroidal direction. We use 64 different field lines, allowing us to perform an ensemble average over 64 samples. It also naturally provides us uncertainties as we can create a distribution function with 64 samples. In doing so, we assume that the statistical properties of perturbed quantities on a flux surface are same while the instantaneous realization of perturbed quantities on different field lines are independent to each other. Correlation functions and their envelopes (using the Hilbert transform) are generated, and we define the correlation length to be the half width at half maximum (HWHM)\footnote{Some may argue that HWHM is not standard as a correlation length. However, we state that our conclusions based on our proposed technique do not rely on absolute size of correlation lengths, rather they depend on either ratios of two different correlation lengths or temporal transition of correlation lengths. Therefore, our conclusions do not change as long as we consistently employ the same statistical technique in this work.} of the envelope of the correlation function in this work.

We provide a detailed description of our method how we obtain correlation functions and correlation lengths. All the reported correlation lengths in this work are estimated at the location of the highest equilibrium pressure gradient. As the perturbed (fluctuating) quantities such as the pressure $P_1$ and the radial component of the magnetic fields $B_{1r}$ are calculated within the simulation separately from the equilibrium ones, i.e., $P_0$ and $B_0$, we do not need to remove `mean' quantities from the numerical data to calculate correlation functions. Note that such a `mean' quantity must be removed first when dealing with experimentally obtained data as they are likely to contain both equilibrium (mean) and perturbed quantities. 

Our data are generated on a field-aligned coordinate system. To explain the coordinate system with an example, let us denote a spatial position $\vec r$ as $\vec r = \psi\hat x + \theta\hat y + (\phi - \phi_\text{shift})\hat z$, where $\psi$ is the usual flux-surface label with $\hat x$ being in the radial direction. $\theta$ denotes the poloidal location while $\hat y$ is in the parallel direction with respect to the background magnetic field. $\phi$ stands for the toroidal position, while $\phi_{\text{shift}}(\psi, \theta) = \int_{\theta_0}^{\theta} \nu(\psi, \theta) d\theta$ with $\hat z$ being in the toroidal direction. Here, $\nu(\psi, \theta)$ is the local field line pitch.\cite{Dudson:2009ig, Leddy:2017jyba} In this coordinate system, spatial positions with the same $\psi$ and the same $\phi-\phi_\text{shift}$ are within a same magnetic field line.

To generate a correlation function, we first build a three-dimensional array of data $D[i_s, j_t, k_e]$ from the simulation data set, where three indices $i_s$, $j_t$ and $k_e$ represent \textit{s}pace (radial, poloidal or parallel), \textit{t}ime and \textit{e}nsemble, respectively. To get a radial correlation function, for instance, we select the data as a function of $\psi$ at a fixed time, and fixed $\theta$ and $\phi-\phi_\text{shift}$ providing us an one-dimensional data set, i.e., $D[i_s, j_0, k_0]$, where $j_0$ and $k_0$ represent the selected time and a specific magnetic field line $\phi-\phi_\text{shift}$ while $i_s$ now contains the radial direction information. By retrieving data from different time and the magnetic field lines, we obtain a set of three-dimensional data $D[i_s, j_t, k_e]$. We, then, obtain a correlation function by shifting the index $i_s$ and performing an ensemble average over $k_e$, which is a radial correlation function, in this example, as a function of time $j_t$. Likewise, to obtain poloidal and parallel correlation functions, we change our choice of $i_s$ so that the data set $D[i_s, j_t, k_e]$ contains poloidal and parallel information.

By forcing data points in $i_s$ to be evenly spaced (see \refapp{app:interpolation_method}) with the distance of neighboring data points $\Delta_s$, we calculate a correlation function $\mathcal{C}(\iota\Delta_s)$ as
\begin{align}
\mathcal{C}\lp \iota \Delta_s \rp [j_t, k_e] = \frac{1}{n-\iota} \sum_{i_s=1}^{i_s=n-\iota} & D[i_s,j_t,k_e] \nonumber \\
\times & D[i_s+\iota, j_t, k_e],
\end{align}
where $\iota$ is the `lag' index, and $n$ is the total number of data points in $i_s$. If the estimated correlation function exhibits a wave-like structure, then we find the envelope of the function using the well-known Hilbert transform:
\begin{equation}
\mathcal{C}_\text{env} \lp \iota \Delta_s \rp = \sqrt{\mathcal{C}^2 \lp \iota \Delta_s \rp + \mathcal{H}^2\lsb \mathcal{C} \lp \iota \Delta_s \rp \rsb},
\end{equation}
where $\mathcal{H}[\cdot]$ is the Hilbert transform operator. 

Ensemble average is performed over different magnetic field lines, i.e., over the index $k_e$, by assuming that statistical properties of perturbed quantities over different field lines within a flux-surface are similar while individual realization is independent. We have selected $64$ different field lines for this purpose, and the ensemble averaged correlation function $\bar{\mathcal{C}}_\text{env} $ is, then,
\begin{align}
\label{eq:mean_std_function}
\bar{\mathcal{C}}_\text{env} \lp \iota \Delta_s \rp [j_t] =  \frac{1}{64} \sum_{k_s=1}^{64} \mathcal{C}_\text{env} \lp \iota \Delta_s \rp[j_t, k_s].
\end{align}

\begin{figure}[t]
\includegraphics[width=\linewidth]{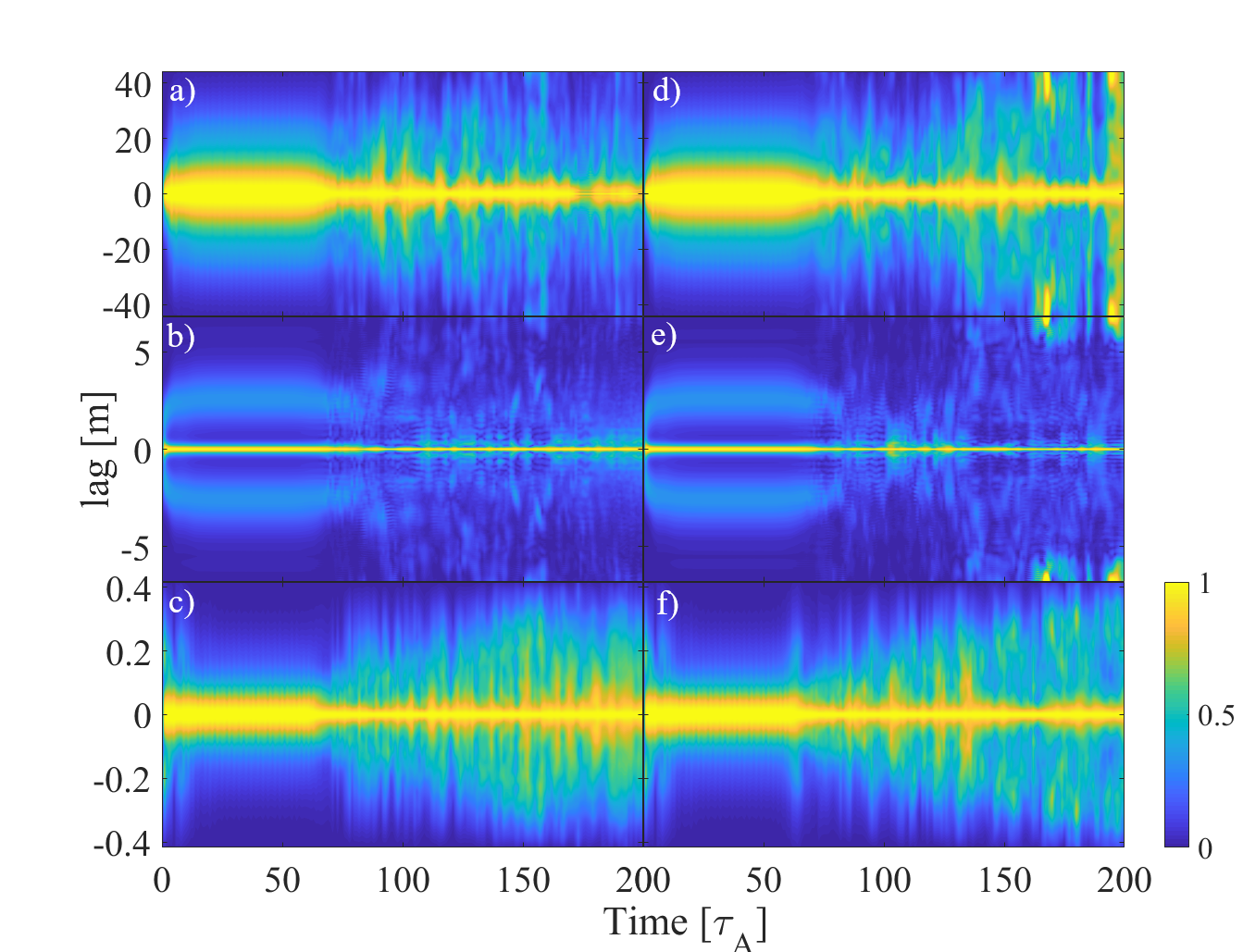}
\caption{Time evolutions of ensemble averaged parallel (top), poloidal (middle) and radial (bottom) correlation functions of $B_{1r}$, i.e., $\bar{\mathcal{C}}_{env} \lp \iota \Delta_s \rp [j_t]$. Left (right) panels are without (with) zonal flows.}
\label{fig:B_correlation_function}
\end{figure}
\reffig{fig:B_correlation_function} shows examples of ensemble averaged parallel (top), poloidal (middle) and radial (bottom) correlation functions as a function of lag distance $\iota\Delta_s$ in the unit of meter and time in the unit of Alfv\'{e}n time $\tau_A$. Left (right) panels are without (with) zonal flows.

\begin{figure}
\includegraphics[width=\linewidth]{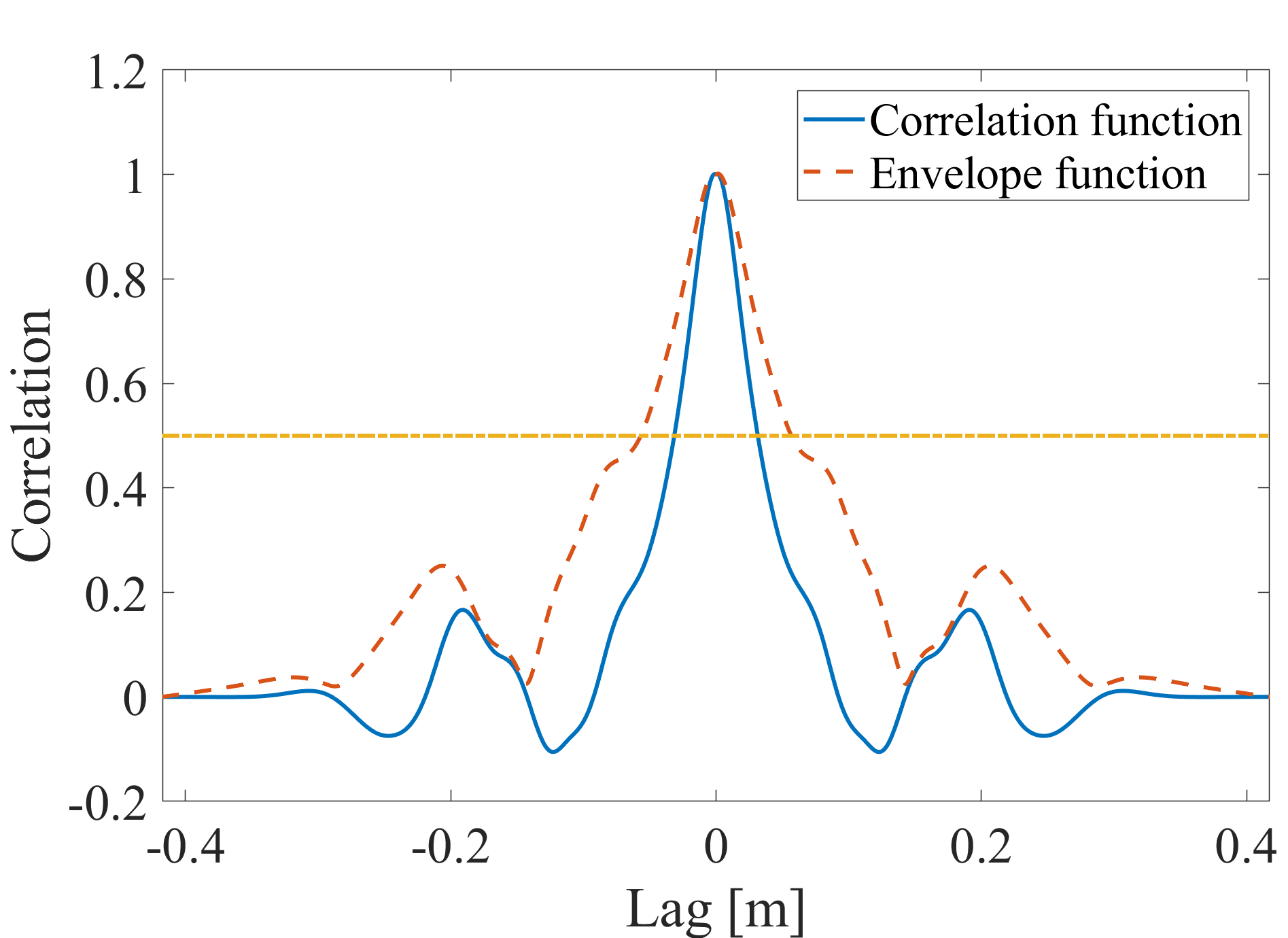}
\caption{An example of the radial correlation function (blue line) and its envelope (red dashed line) at a fixed time before an ensemble average. Yellow horizontal line indicates the correlation value of $0.5$, i.e. half maximum.}
\label{fig:correlation_length_measurement}
\end{figure}

Once we have a correlation function, we obtain the ensemble averaged correlation length $l$ and its uncertainty $\sigma_l$ as following:
\begin{align}
l[j_t] = \frac{1}{64}\sum_{k_s=1}^{64} \text{HWHM}\{\mathcal{C}_\text{env}(\iota\Delta_s)[j_t, k_s]\}, \\
\sigma^2_l[j_t] = \frac{1}{64}\sum_{k_s=1}^{64} \lsb   \text{HWHM}\{\mathcal{C}_\text{env}(\iota\Delta_s)[j_t, k_s]\} - l[j_t]   \rsb^2,
\end{align}
where $\text{HWHM}\{\cdot\}$ returns the half-width at half-maximum.\footnotemark[\value{footnote}] \reffig{fig:correlation_length_measurement} is an example of radial correlation function (blue line) and its envelope (red dashed line) at a fixed time ($\tau_A=99$) before the ensemble average. $64$ of such functions are used to estimate the ensemble averaged radial correlation length and its uncertainty. 

All the reported radial, poloidal and parallel correlation lengths and their uncertainties in this work are estimated based on the described procedure in this section. 

\begin{figure}[t]
\includegraphics[width=\linewidth]{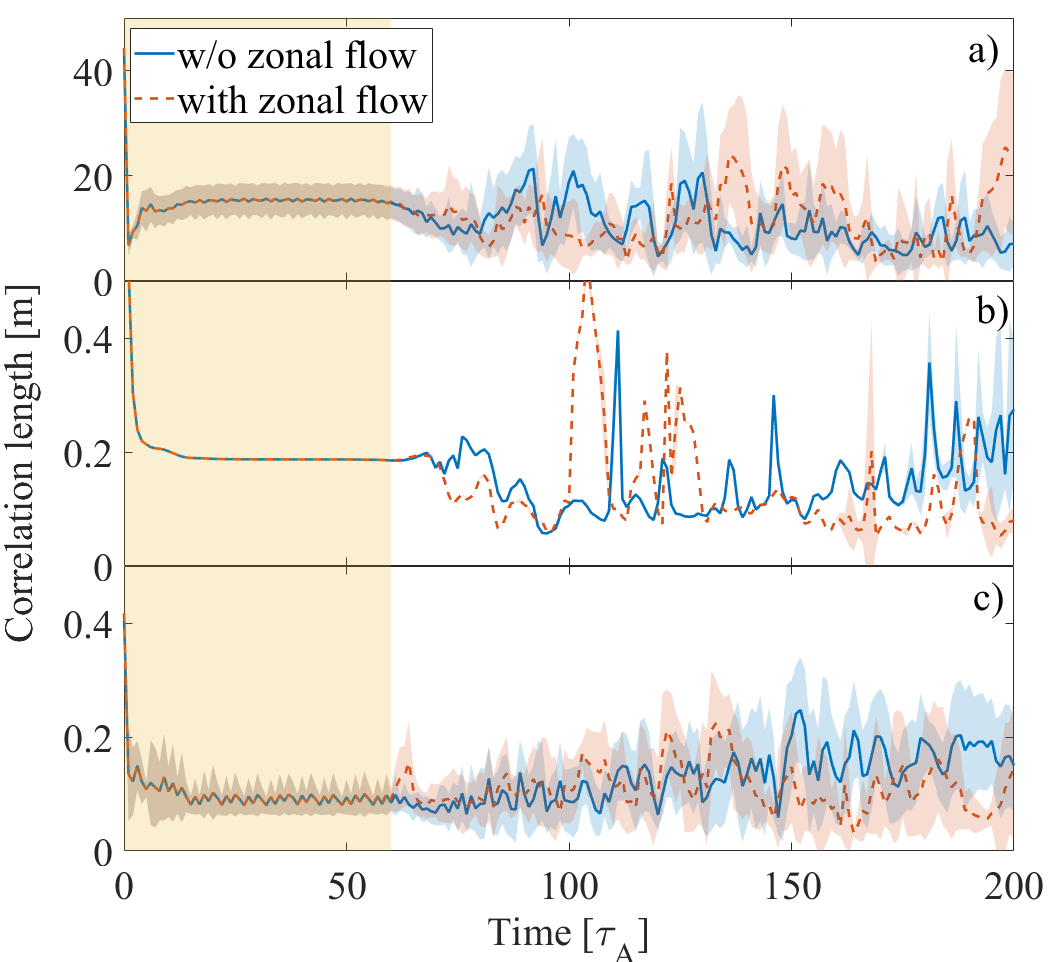}
\caption{Temporal evolutions of (a) parallel, (b) poloidal and (c) radial correlation lengths of radial component of the perturbed magnetic fields $B_{1r}$ with (red dashed line) and without (blue line) zonal flows. Shades around the lines represent the uncertainties of the correlation lengths. $\tau_A<60$ (yellow shaded area) is the linear state.}
\label{fig:B_correlation_length}
\end{figure}

We have estimated parallel, poloidal and radial correlation lengths of $B_{1r}$ based on the corresponding correlation functions shown in \reffig{fig:B_correlation_function}. \reffig{fig:B_correlation_length} shows temporal evolutions of (a) parallel, (b) poloidal and (c) radial correlation lengths with (red dashed line) and without (blue line) zonal flows. During the nonlinear state (and stochastic region) parallel correlation lengths are typically $10-20$\:m; while poloidal and radial correlation lengths are about a factor of ten smaller. $\tau_A<60$ (yellow shaded area) is the linear state, therefore the correlation lengths in this area should not be given any significant physical meaning.

\section{Magnetic Kubo number and correlation length of pressure perturbations}
\label{sec:kubo_number}

For the sake of the readers, we provide the definition of the magnetic Kubo number, i.e., \refeq{eq:kubo_number}, once again here:
\begin{equation}
\mathcal{R}= \frac{ \delta B }{B_0} \frac{l_\parallel}{l_\perp}, \nonumber
\end{equation}
which states that estimation of $\mathcal{R}$ requires information of the fluctuation level and correlation lengths in parallel and perpendicular directions (with respect to the equilibrium magnetic field) of stochastic magnetic fields along with the knowledge of equilibrium magnetic fields.

\begin{figure}[t]
\includegraphics[width=\linewidth]{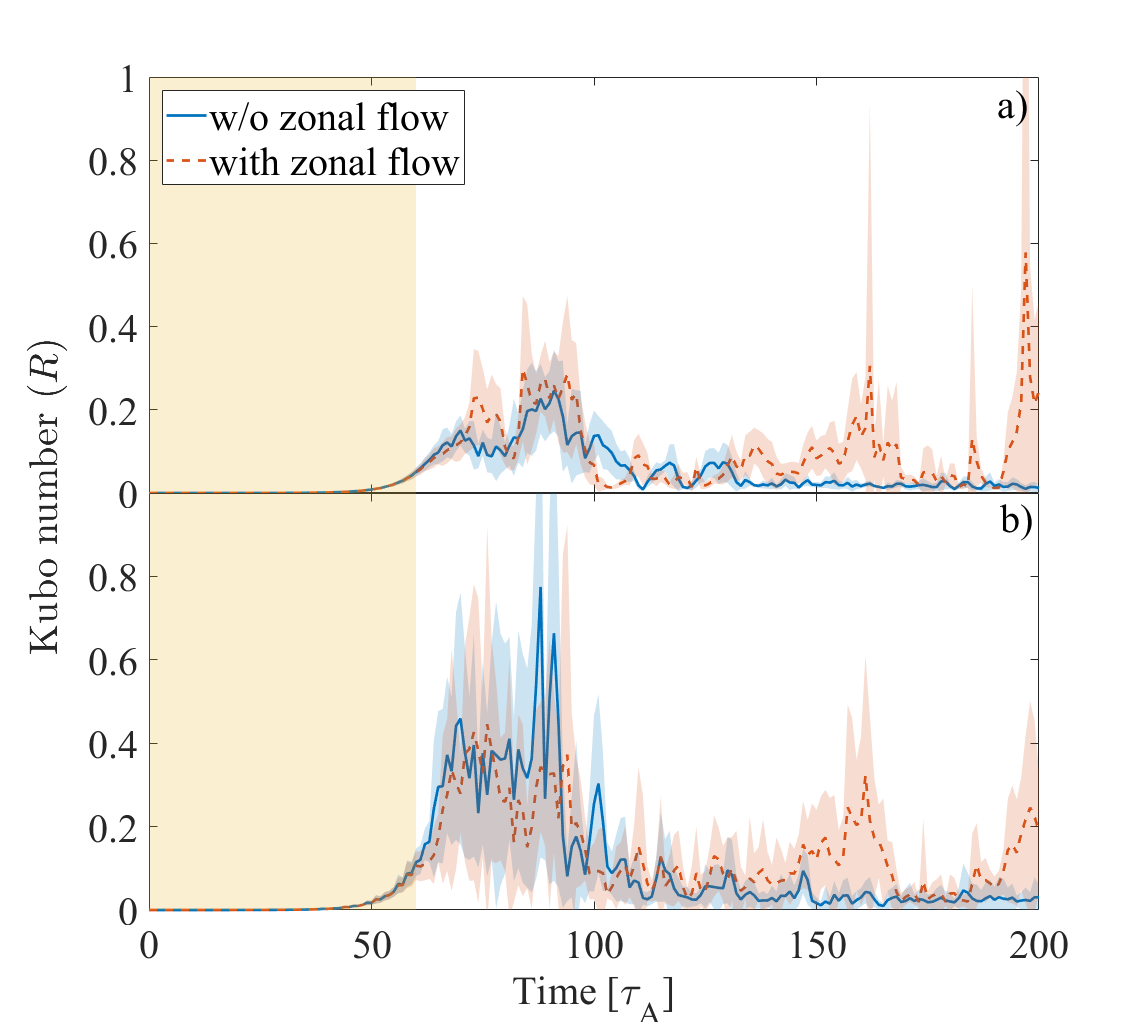}
\caption{Temporal evolutions of the magnetic Kubo number using (a) poloidal and (b) radial correlation length as a measure for the perpendicular one, $l_\perp$ in \refeq{eq:kubo_number} with (red dashed line) and without (blue line) zonal flows. Kubo number is found to be always less than one during the edge pedestal collapse. Shades around the lines represent the uncertainties of the Kubo number. $\tau_A<60$ (yellow shaded area) is the linear state.}
\label{fig:kubo_number_graph}
\end{figure}

We estimate the fluctuation level as a spatial (poloidal and toroidal) standard deviation of $B_{1r}$ on the flux surface where $\alpha=3.87$. As for the perpendicular correlation lengths, we have two options: radial vs. poloidal correlation lengths. Even though the \textit{radial} correlation length will be more relevant as it is associated with the \textit{radial} cross-field diffusion process in magnetically confined plasmas, we provide the temporal evolution of the Kubo number with (a) poloidal and (b) radial correlation lengths with (red dashed line) and without (blue line) zonal flows in \reffig{fig:kubo_number_graph}. The linear state ($\tau_A<60$) is shaded in yellow. 

\reffig{fig:kubo_number_graph} is the key result of this paper. One can immediately find that the Kubo number $\mathcal{R}$ is always less than one regardless of using radial or poloidal correlation lengths as a measure for perpendicular correlation, $l_\perp$ in evaluating the Kubo number. Also, it does not depend much on the presence of zonal flows. In more detail, we have $0.2<\mathcal{R}<1.0$ when the Chirikov parameter $\mathcal{C}$ is much larger than unity shown in \reffig{fig:chirikov_parameter}, i.e., $60\lesssim\tau_A\lesssim 80$; whereas $\mathcal{R}\lesssim 0.2$ when $\mathcal{C}\approx 1$, i.e., $\tau_A\gtrsim 80$. Therefore, our result suggests that during the abrupt edge pedestal collapse ($60\lesssim\tau_A\lesssim 80$) the cross-field diffusion could be explained by the quasilinear Gaussian diffusion process.\cite{Isichenko:1991ec, Pommois:2001cl,Pommois:1999ei,Pommois:1998fzba, Zimbardo:2000cv, Zimbardo:2000fm} The fact that $\mathcal{R}<1$ in all the time implies that the percolation theory may not be applicable to describe the radial transport process during an edge pedestal collapse.\cite{Kadomtsev:1979} This justifies, in some sense, the usage of the modified Rechester-Rosenbluth model\cite{Rechester:1978cl} described in \refsec{sec:nume_model}. Note that the initial pressure profile being used in simulations is highly unstable to the ideal ballooning mode. Therefore, in actual experiments where $\alpha \gtrsim \alpha_c$ is satisfied, the strength of stochastic fields will be weaker than the  simulation result. This implies that the actual Kubo number will be smaller than that given in \reffig{fig:kubo_number_graph}.

Concerning actual experiments in magnetic fusion plasmas, it is difficult to have a direct measurement of the magnetic Kubo number, if not impossible. Such a task would require performing the fine (probably ion-scale) structure current tomography based on measurements of magnetic fields and fluxes. There exists a high performance current tomography technique based on a Bayesian approach,\cite{Svensson:2008inba} but it is not as fine as the required ion-scale tomography. On the contrary, diagnostic systems for ion-scale fluctuations of density or temperature are well developed. For instance, KSTAR is equipped with beam emission spectroscopy (BES)\cite{Nam:2014kdba, Nam:2012caba}, microwave imaging reflectometry (MIR)\cite{Lee:2014jwba} and electron cyclotron emission imaging (ECEI)\cite{Yun:2010esba} which are capable of measuring such fluctuations of density or temperature. Many other magnetic confinement devices do have similar diagnostic systems. Thus, we provide characteristics of pressure fluctuations observed in the simulation which may potentially be used as a circumstantial (or indirect) evidence for the onset of stochastic magnetic fields, if similar characteristics are observed in real experiments during an edge pedestal collapse.

\begin{figure}
\includegraphics[width=\linewidth]{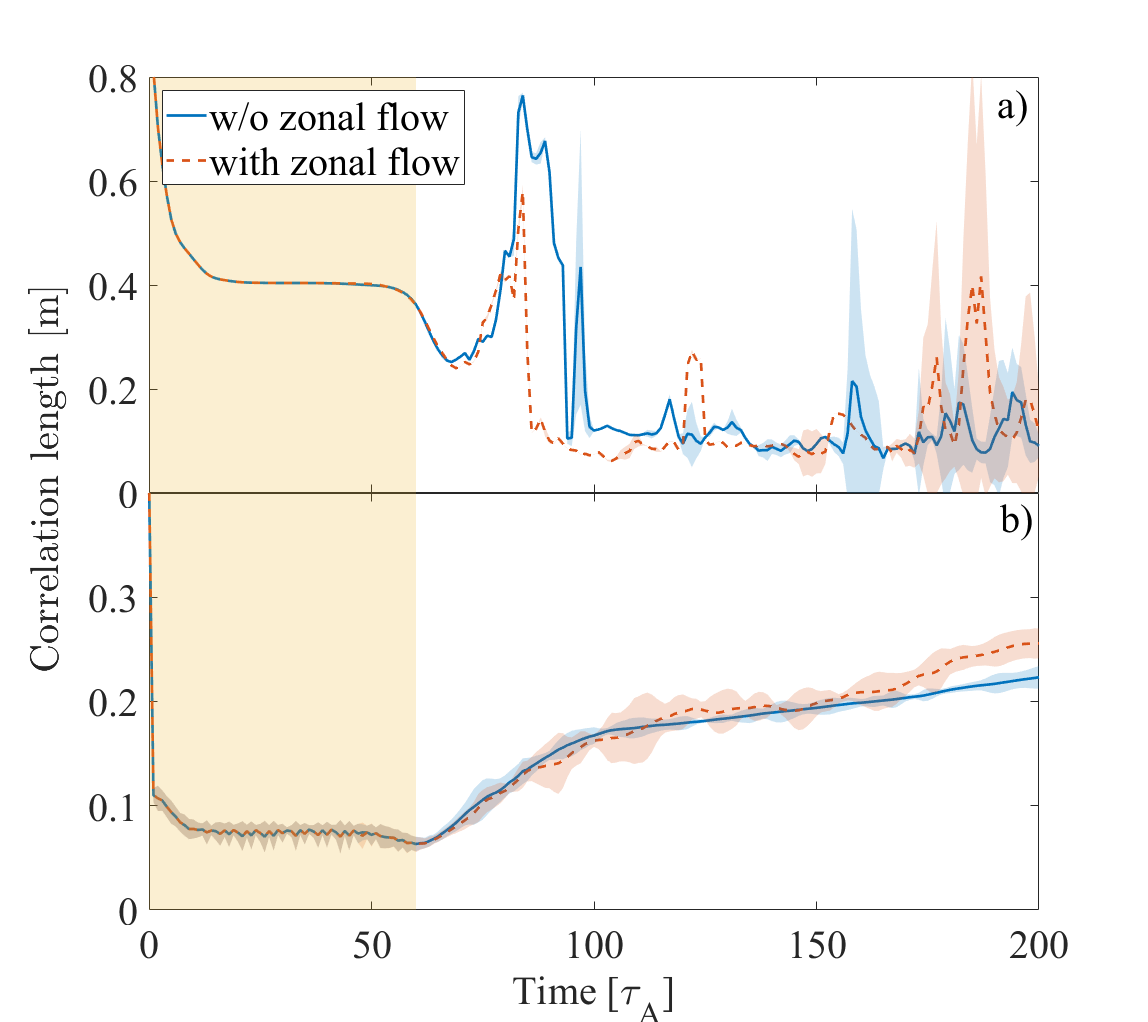}
\caption{(a) Poloidal and (b) radial correlation lengths with (red dashed line) and without (blue line) zonal flows. Temporal evolutions of the poloidal correlation length resembles those of the Chirikov parameter and the Kubo number. Evolution of the radial correlation length could be explained by the widening of the stochastic layer width. $\tau_A<60$ (yellow shaded area) is the linear state.}
\label{fig:P_correlation_length}
\end{figure}

We have filtered out the $(m,n)=(1,0)$ component of pressure fluctuations as it corresponds to the geodesic acoustic mode (GAM)\cite{Winsor:1968jpbaca} when we perform correlation analyses on our simulation data. \reffig{fig:P_correlation_length} shows (a) poloidal and (b) radial correlation lengths with (red dashed line) and without (blue line) zonal flows. The poloidal correlation length increases suddenly and falls down to a certain plateau value whose behavior resembles the temporal evolutions of Chirikov parameter (\reffig{fig:chirikov_parameter}) and the Kubo number (\reffig{fig:kubo_number_graph}). 
\begin{figure}
\includegraphics[width=\linewidth]{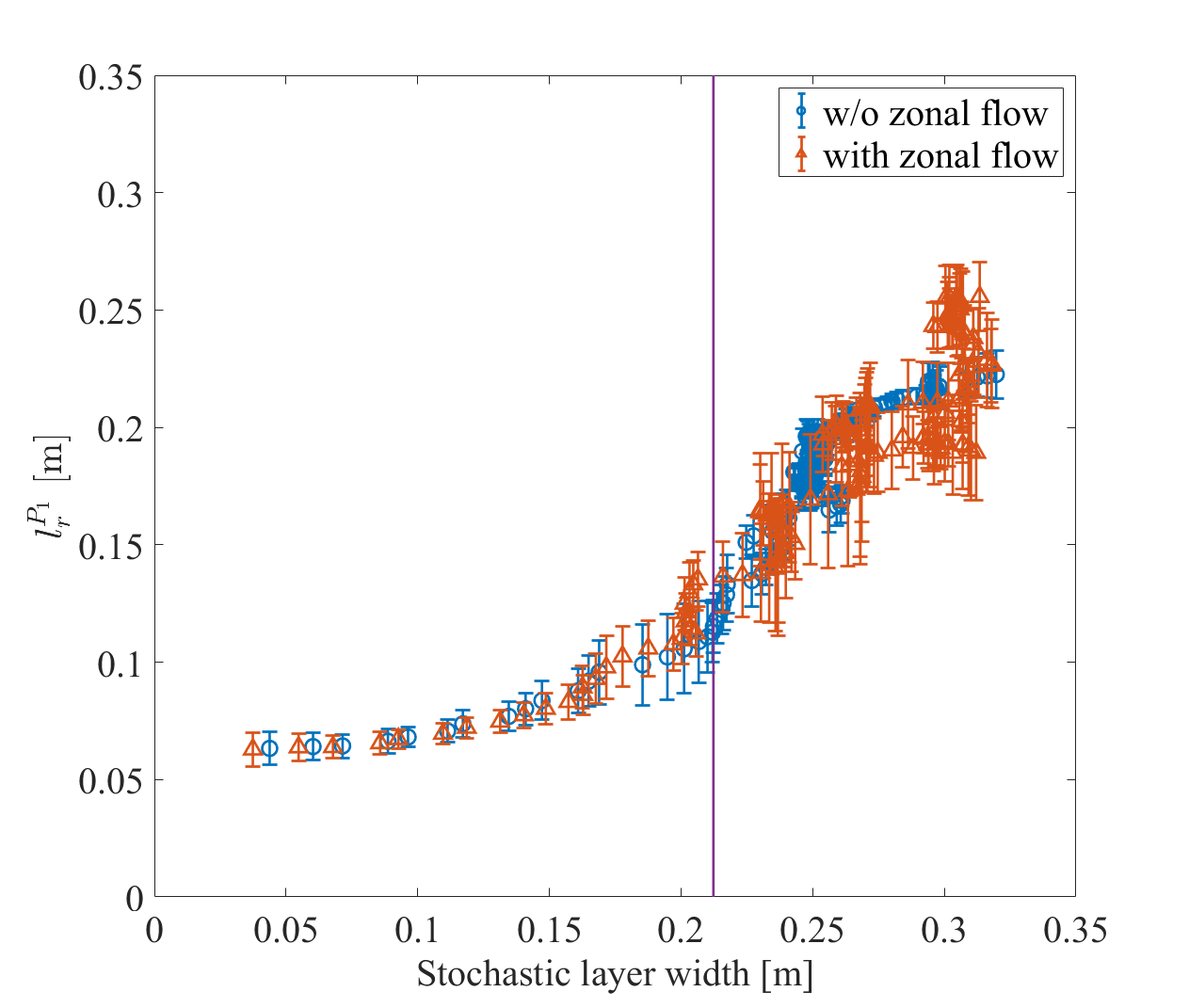}
\caption{Radial correlation lengths $l_r^{P_1}$ vs. stochastic layer width with (red) and without (blue) zonal flows showing that they are well correlated to each other. Data points on the left of the purple vertical line correspond to $l_r^{P_1}$ during the abrupt edge pedestal collapse ($60\lesssim\tau_A\lesssim 80$), and those on the right part are taken after the collapse ($\tau_A \gtrsim 80$).}
\label{fig:correlation_vs_width}
\end{figure}

The radial correlation length slowly increases which may be explained by the widening of the stochastic layer width shown in \reffig{fig:stochastic_layer}. \reffig{fig:correlation_vs_width} clearly shows such a correlation between the two. We find that during the abrupt edge pedestal collapse ($60\lesssim\tau_A\lesssim 80$) stochastic layer width increases faster than that of the radial correlation length (left part of the purple vertical line), whereas once the collapse is completed ($\tau_A \gtrsim 80$) their growth rates become comparable. 

In actual experiments, the poloidal and radial correlation lengths of density fluctuations can be easily measured by an appropriate arrangement of the diagnostic devices. Therefore, we suggest that a sudden drop of poloidal correlation lengths of density fluctuation accompanied with the increase of radial correlation length could be an indirect indicator of the magnetic field stochastization during an pedestal collapse.

\section{Conclusion}
\label{sec:conclusion}

We perform a correlation analysis to find out how the magnetic Kubo number $\mathcal{R}$ evolves during and after an edge pedestal collapse. Results of numerical simulation reported in \citet{Rhee:2015czbaca} and \citet{Jhang:2017jm} are used in the analysis. Our analysis shows that $\mathcal{R}$ never exceeds the unity even if we have initiated the simulation with a highly unstable plasma. $\mathcal{R}$ is found to be within $0.2$ and $0.6$ during the collapse, while it becomes less than $0.1$ once the collapse is completed. Based on our results, we suggest that the quasilinear Gaussian diffusion process could be adequate to model the cross-field diffusion process during the collapse. In all cases, the percolation theory may not be necessary to describe the transport process during and after the collapse. We note that the effect of zonal flows on Kubo number is minimal in this work.

Since experimental identification of stochastic magnetic fields and the estimation of Kubo number are challenging tasks, we have discussed how pressure fluctuations evolve during and after the edge pedestal collapse. Ion--scale pressure fluctuation measurements are routinely available in modern magnetic confinement devices. We find that the poloidal correlation length abruptly increases during the collapse, drops quickly to a lower value and stays there whose temporal evolution is close to that of both Chirikov parameter and Kubo number. The radial correlation length is found to be strongly correlated with the stochastic layer width.  If such behavior is observed in experiments, then it may be used as a circumstantial evidence of stochastic magnetic fields. This is planned as a future work.

\begin{acknowledgments}
We are grateful to Dr. T. Rhee and Dr. Juyoung Kim for informative discussions. Hogun Jhang conceived this problem during the stimulating discussion at the 5th APTWG meeting (Dalian, China). This research was mainly supported by the R\&D Program through National Fusion Research Institute funded by the Ministry of Science and ICT of the Republic of Korea(NFRI-EN1741-3). WL and YCG are partially supported by National R\&D program through the National Research Foundation of Korea (NRF) funded by the Ministry of Science and ICT of the Republic of Korea (grant number NRF-2017M1A7A1A01015892) and the KUSTAR-KAIST Institute, KAIST, Korea.
\end{acknowledgments}

\appendix
\counterwithin{figure}{section}

\section{A method to obtain evenly spaced data points in BOUT++}
\label{app:interpolation_method}

\begin{figure}
\includegraphics[width=\linewidth]{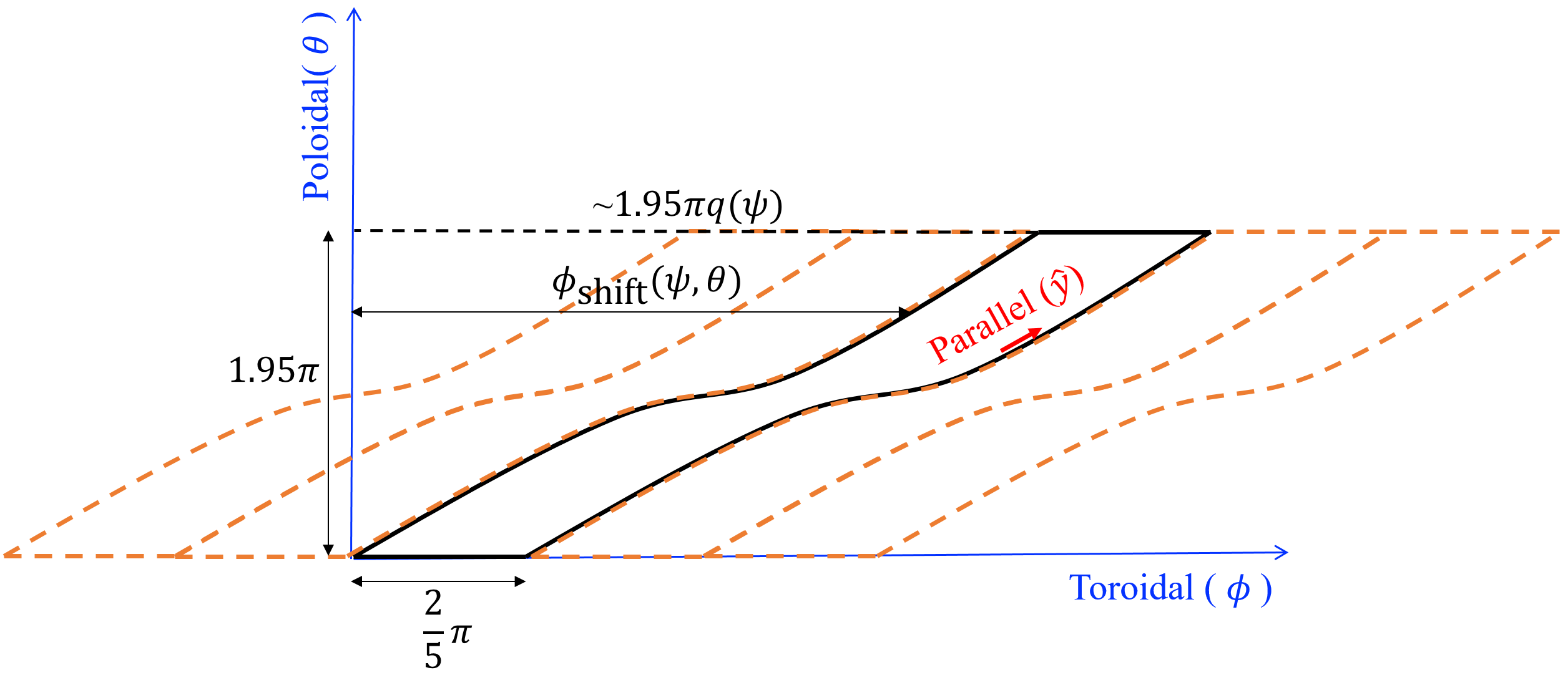}
\caption{A cartoon picture of the simulation domain (black box) in a $\theta-\phi$ flux surface coordinate. Yellow dashed boxes show periodic boundary conditions in toroidal directions resulting from the Fourier transform in toroidal direction.}
\label{fig:extrapolation_fig}
\end{figure}

We have described the coordinate system of the BOUT++ framework in \refsec{sec:mag_corr_length_section}, where a spatial position $\vec r$ is written as $\vec r = \psi\hat x + \theta\hat y + (\phi - \phi_\text{shift})\hat z$ with $\hat x$, $\hat y$ and $\hat z$ denoting radial, parallel and toroidal directions, respectively. For clarity and simplicity, let us define new variables here: $x_i=\psi_i$, $y_j=\theta_j$ and $z_k=\phi_k - \phi_\text{shift}(\psi_i, \theta_j)$ where subscripts $i$, $j$ and $k$ select the spatial position of interest. The domains of the simulation used in this work are $-0.4824<x<0.2573$, $0<y<1.95\pi$ and $0<z<2\pi/5$ with $516 (i) \times64 (j) \times64 (k)$ data points. A black box in \reffig{fig:extrapolation_fig} shows the simulation domain.

By extracting the data with fixed values of $x_i$ and $z_k$ with varying $y_j$, we obtain a data set from a certain equilibrium field line allowing us to estimate a parallel correlation length; while estimating radial and poloidal correlation lengths require data sets from a fixed toroidal position $\phi_k=z_k + \phi_\text{shift}(x_i, y_j)$ with varying $x_i$ or $y_j$.  Since $\phi_\text{shift}(x_i, y_j)$ changes its value as we vary $x_i$ or $y_j$, it is difficult to fix the value of $\phi_k$ in the original simulation domain. For this reason, we represent the numerical data in the Fourier domain for $z$-component, resulting in a periodic boundary condition in the toroidal direction as shown in \reffig{fig:extrapolation_fig} with yellow dashed boxes.

The original BOUT++ data, $s(x_i,y_j, z_k)$, in $z$ direction are evenly spaced with the spatial resolution of $\Delta_z=z_{k+1}-z_k=\pi/160$, i.e., 64 data points covering $2\pi/5$ radians, and its Fourier representation with the corresponding data points $N=64$ can be written as
\begin{eqnarray}
\hat S(x_i,y_j, \kappa) &=& \sum_{k=0}^{63} s(x_i,y_j, z_k) e^{-i \kappa \frac{2\pi}{N} \frac{z_k}{\Delta_z}}\nonumber \\ 
&=& \sum_{k=0}^{63} s(x_i,y_j, z_k)  e^{-i\kappa 5 z_k},
\end{eqnarray}
where $\kappa$ is the index of the toroidal wavenumber. Then, to obtain a data set with a fixed value of $\phi_k$, for instance $\phi_k=\phi_0$, we perform an inverse Fourier transform as
\begin{eqnarray}
s(x_i,y_j, z(\phi_0)) &=& \frac{1}{64}\sum_{\kappa=0}^{63} \hat S (x_i,y_j, \kappa) e^{i \kappa \frac{2\pi}{N} \frac{\phi_0-\phi_\text{shift}(x_i, y_j)}{\Delta_z}} \nonumber \\ 
&=& \frac{1}{64} \sum_{\kappa=0}^{63} \hat S(x_i,y_j, \kappa)  e^{i \kappa 5 (\phi_0-\phi_\text{shift}(x_i, y_j))}. \nonumber \\
\end{eqnarray}
Thus, if we wish to obtain a radial correlation function at a fixed poloidal ($y_j = y_0$) and toroidal ($\phi_k=\phi_0$) position, we use a data set $s(x, y_0, z(\phi_0))$; or to obtain a poloidal correlation function at a fixed radial ($x_i=x_0$) and toroidal ($\phi_k=\phi_0$) position, we use a data set $s(x_0, y, z(\phi_0))$. For ensemble averaging of the correlation functions described in \refsec{sec:mag_corr_length_section}, we obtain 64 samples from different toroidal locations of $\phi=0,\pi/32,2\pi/32,...,63\pi/32$ on a flux surface.

As the $(x, y)$ coordinate with a fixed $\phi=\phi_0$ can be readily transformed into a cylindrical coordinate system $R(x, y; \phi_0)$ and $Z(x, y; \phi_0)$ in our numerical simulation data, we can approximate the distance between the neighboring spatial data points to
\begin{align}
L^{r}_{i,j \rightarrow i+1, j} \approx& \{[R(x_i,y_j; \phi_0)-R(x_{i+1},y_{j}; \phi_0)]^2 \nonumber \\
+&[Z(x_i,y_j; \phi_0)-Z(x_{i+1},y_{j}; \phi_0)]^2\}^{1/2}
\end{align}
where $L^{r}_{i,j \rightarrow i+1, j}$ denotes the radial distance between the two points of $(R(x_i, y_j; \phi_0), Z(x_i, y_j; \phi_0))$ and $(R(x_{i+1}, y_j; \phi_0), Z(x_{i+1}, y_j; \phi_0))$. Then, the distance between any two radially separated spatial positions, i.e., between $(R(x_{i_1}, y_{j_1}; \phi_0), Z(x_{i_1}, y_{j_1}; \phi_0))$ and $(R(x_{i_2}, y_{j_1}; \phi_0), Z(x_{i_2}, y_{j_1}; \phi_0))$,  can be estimated as a sum of the distance between the neighboring points from the start to the end positions:
\begin{equation}
\label{eq:distance_sum_eq}
L^{r}_{i_1, j_1 \rightarrow i_2, j_1} \approx \sum_{i=i_1}^{i=i_2-1} L^{r}_{i , j_1 \rightarrow i+1,j_1}.
\end{equation}
Estimation of the poloidal separation distance can be done similarly. A parallel distance between the two neighboring spatial positions, i.e. between $(x_i, y_j; z_0)$ and $(x_i, y_{j+1}; z_0)$ with a fixed field line $z_0$, can be estimated as 
\begin{eqnarray}
L^{\parallel}_{i,j\rightarrow i,j+1} &\approx& [\phi_{shift}(x_i, y_j; z_0)-\phi_{shift}(x_{i},y_{j+1}; z_0)] \nonumber \\
&&\times R(x_i, y_j; z_0). 
\end{eqnarray}
As for the case of radial and poloidal distance between any two arbitrary positions, we estimate such parallel distance by summing over all the neighboring distances from the start to the end positions. Once we have all the distance information from one position to another, we use an interpolation scheme to generate evenly spaced data sets to calculate correlation functions as described in \refsec{sec:mag_corr_length_section}.

For radial data sets, we have $0.058$\:cm as a spatial resolution with the radial domain of $\sim[4.39, 4.81]$\:m. For poloidal (parallel) data sets, we have a spatial resolution of $0.67$ ($35$)\:cm with the domain size of $6.81$ ($44.3$)\:m. The reference position for all the correlation functions is set to be at $R=4.59$\:m and $Z=0.047$\:m corresponding to the position of the largest pressure gradient with approximately $5$\:cm away from the midplane.

\bibliographystyle{apsrev4-1}
\bibliography{pop2017_ref}

\end{document}